# Rashba effect modulation in two-dimensional $A_2B_2Te_6$ (A = Sb, Bi; B = Si, Ge) materials via charge transfer

Haipeng Wu[1], Qikun Tian[2], Jinghui Wei[1], Ziyu Xing[1], Guangzhao Qin[2], Zhenzhen Qin[1,−]

[1]Key Laboratory of Materials Physics, Ministry of Education, School of Physics, Zhengzhou University, Zhengzhou 450001, P. R. China

[2]State Key Laboratory of Advanced Design and Manufacturing Technology for Vehicle, College of Mechanical and Vehicle Engineering, Hunan University, Changsha 410082, P. R. China

## Abstract

Designing two-dimensional (2D) Rashba semiconductors, exploring the underlying mechanism of Rashba effect, and further proposing efficient and controllable approaches are crucial for the development of spintronics. On the basis of first-principles calculations, we here theoretically design all possible types (common, inverse, and composite) of Janus structures and successfully achieve numerous ideal 2D Rashba semiconductors from a series of five atomic-layer $A_2B_2Te_6$ (A = Sb, Bi; B = Si, Ge) materials. Considering the different Rashba constant $\alpha_R$ and its modulation trend under external electric field, we comprehensively analyze the intrinsic electric field $E_{in}$ in terms of work function, electrostatic potential, dipole moment, and inner charge transfer. Inspired by the quantitative relationship between charge transfer and the strength of $E_{in}$ and even the $\alpha_R$, we propose a straightforward strategy of introducing a single adatom onto the surface of 2D monolayer to introduce and modulate the Rashba effect. Lastly, we also examine the growth feasibility and electronic structures of the Janus $Sb_2Ge_2Se_3Te_3$ system and Janus-adsorbed systems on a 2D BN substrate. Our work not only conducts a detailed analysis of $A_2B_2Te_6$-based Rashba systems, but also proposes a new strategy for efficiently and controllably modulating the $\alpha_R$ through the reconfiguration of charge transfer.

− Corresponding author: qzz@zzu.edu.cn

# I. Introduction

Spin-orbit coupling (SOC) is the interaction between the spin and orbital motion of electrons, originating from the effective magnetic field experienced by electrons in crystals within their motion framework[1,2]. The Rashba-type SOC exists in the spin splitting of two-dimensional (2D) electronic states and is particularly common in Janus structures, crystal surfaces or interfaces where the inversion symmetry is broken, as it can manipulate electron spin without the need for an external magnetic field, making it possible to be utilized in the next generation of spintronic devices, such as spin field effect transistors (SEFTs)[3–6].

Since 2017, when the milestone synthesis of MoSSe as the first Janus TMD[7], the 2D Janus structures with intrinsic symmetry breaking have attracted great attention as a good carrier for achieving ideal Rashba effect[8,9]. Designing Janus structures with two- and three-atom layers is relatively common and can be achieved by substituting the outermost atoms[10–14], such as PX (X = As, Sb, Bi)[12] and MXY (M = Mo, W; X, Y = S, Se, Te; X ≠ Y)[13]. In reality, polyatomic layer structures provide richer choices for Janus structural design, such as four-atom layer structures $X_2Y_2$, which can be designed as $XSn_2Y$[15], $XYZ_2$[16], and ZnAXY[17] based on the outer, inner, and double-layer substitution. Rashba effect can be described by the Bychkov-Rashba Hamiltonian form: $H_R = \alpha_R (\sigma \times k) \cdot z$, where the $\alpha_R$ represents the Rashba constant, $\sigma$ denotes the Pauli spin matrices, $k$ stands for the momentum, and $z$ is the electric field direction[18]. The $\alpha_R$ can be modulated by various external strategies, including electric fields[19–22], strains[23,24]. While the variation of the strength of the Rashba effect under external electric fields ($E_{ex}$) is almost always linear, the results of $\alpha_R$ decreasing or increasing with the $E_{ex}$ are unpredictable. For instance, a positive electric field decreases $\alpha_R$ in the MgTe monolayer but increases $\alpha_R$ in the ZnTe and CdTe monolayers[21]. Although the strains may yield more desirable results, the results are often irregular. For example, in the Sb/InSe heterostructure ($\alpha_R$ =1.04 eVÅ), when applying in-plane tensile strain, $\alpha_R$ decreases to 0.9 eVÅ at 2% and then increases to 1.25 eVÅ at 4%[24]. Therefore, it is urgently desired to confirm the key factor that affects $\alpha_R$ and its modulation, and further to seek a simple strategy to tune the Rashba spin splitting (RSS), which is currently needed to design nanoscale structures suitable for spintronic devices.

Recently, $A_2B_2Te_6$ monolayers (such as $Sb_2Si_2Te_6$[25], $Sb_2Ge_2Te_6$[26], and $Bi_2Si_2Te_6$[27]) have been theoretically predicted to be stable. Considering the properties of strong SOC and the multiple atomic layers, constructing 2D Janus structures by breaking the mirror symmetry of such systems is an effective method to introduce the Rashba effect. In this work, we employ first-principles calculations to design various possible types of Janus structures and successfully achieve numerous ideal 2D Rashba semiconductors from a series of five atomic-layer $A_2B_2Te_6$ (A=Sb, Bi; B=Si, Ge) materials. In perspective of Rashba constant $α_R$ and its modulation trend under external eletric field, we comprehensively analyze the intrinsic electric field $E_{in}$ in terms of work function and electrostatic potential, dipole moment, or inner charge transfer. Inspired by the quantitative relationship between charge transfer and the strength of $E_{in}$ and even the $α_R$, we propose a straightforward strategy of introducing a single adatom onto the surface of 2D monolayer to introduce and modulate the Rashba effect. Lastly, we also examine the growth feasibility and electronic structures of the Janus $Sb_2Ge_2Se_3Te_3$ system and Janus-adsorbed systems on a 2D BN substrate.

## II. Computational details

The first-principles calculations were implemented in the Vienna Ab initio Simulation Package (VASP) based on density functional theory (DFT) and the projected-augmented wave (PAW) method[28,29]. The Perdew-Burke-Ernzerhof (PBE) exchange-correlation functional within the generalized gradient approximation was utilized[30]. Both with and without SOC were contained to discuss the phenomena of the Rashba effect. Due to the presence of dipole moment in the Janus structures, the dipole correction was added. During the calculations, the kinetic energy cutoff for the plane wave basis and the $k$-point mesh of the Brillouin zone were set to 500 eV and 12 × 12 × 1, respectively[31]. The tolerance of the total energy was set to be $10^{-6}$ eV with forces less than 0.001 eV/Å. A vacuum layer of 30 Å was used to eliminate unnecessary interaction along $z$ direction between adjacent slabs to ensure that the results were sufficiently accurate. Phonon dispersion analysis was performed with a 5 × 5 × 1 supercell by using the PHONOPY code interfaced with the density-functional perturbation theory[32]. To further confirm the thermal stability of the studied monolayers, the ab initio molecular dynamics (AIMD) simulation was tested at room temperature for 10 ps with a step of 2 fs. In addition, the charge transfer between different layers was analyzed by Bader technique[33]. Atomic structure was visualized by using VESTA software package[34].

## III. Results and discussion

### A. Structures and stabilities of Janus monolayers

Fig. 1(a) illustrates side and top views of monolayers $A_2B_2X_6$ (A = Sb, Bi; B = Si, Ge; X = Te), among

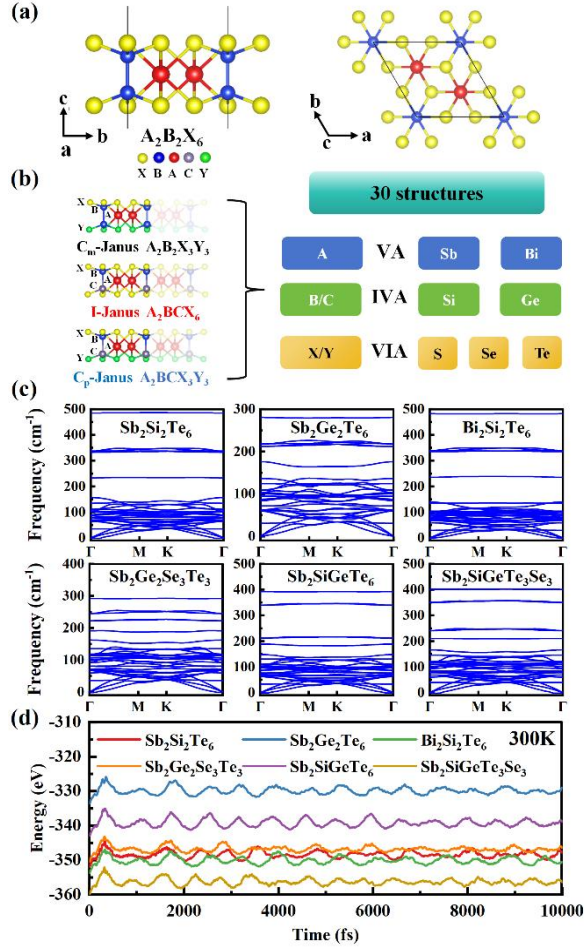

FIG. 1. (a) Side and top views of the crystal structures of the studied original and (b) Janus monolayers. Yellow, blue, red, grey, green balls represent X, B, A, C, Y atoms, respectively. The unit cell is displayed by the solid lines. $C_m$-Janus, I-Janus, $C_p$-Janus represent common, inverse, composite Janus structures, respectively. (c) The phonon spectra and (d) AIMD simulations of $Sb_2Si_2Te_6$, $Sb_2Ge_2Te_6$, $Bi_2Si_2Te_6$, $Sb_2Ge_2Se_3Te_3$, $Sb_2SiGeTe_6$ and $Sb_2SiGeTe_3Se_3$ monolayers.

which X = Te is stemmed from the presence of heavier elements and stronger SOC. The primitive unit cell with space group $P\bar{3}1m$ (No. 162) is formed by ten atoms and indicated by a black parallelogram, including two A atoms, two B atoms, and six X atoms, forming a 2D monolayer with five atomic-layer along the $z$-axis in the order of X, B, A, B, and X. The A atom is located in the center of Te octahedron ($ATe_6$), and $B_2$ dimer with surrounding Te atoms forming $B_2Te_6$ ethane-likes group. Based on the original $A_2B_2Te_6$ structures, 30 types of Janus configurations can be designed by substituting one of outer/ inner atomic layer or both inner and outer layers with different elements, including 12 types of $C_m$-Janus $A_2B_2X_3Y_3$ monolayers (A = Sb, Bi; B = Si, Ge; X, Y = S, Se, Te; X ≠ Y), 6 types of I-Janus $A_2BCX_6$ monolayers (A = Sb, Bi; B = Si; C = Ge; X = S, Se, Te) with the inner layer substitution[16], and 12 types of $C_p$-Janus $A_2BCX_3Y_3$ monolayers (A = Sb, Bi; B = Si; C = Ge; X, Y = S, Se, Te; X ≠ Y), disrupting mirror symmetry and possessing a $P31m$ (No. 157) space group. For instance, the $C_m$-Janus $Sb_2Si_2Se_3Te_3$ monolayer, I-Janus $Si_2SiGeTe_6$ monolayer and $C_p$-Janus

Sb$_2$SiGeSe$_3$Te$_3$ monolayer are constructed by replacing the Te atomic layer with the Se atomic layer in the original Sb$_2$Si$_2$Te$_6$ monolayer, replacing the inner Si atomic layer with the Ge atomic layer, and replacing the Si and Te atomic layers with the Ge and Se atomic layers, respectively [Fig. 1(b)]. Among them, the design of C$_m$-Janus is usual[35,36], and I-Janus has been reported in previous work[16], while C$_p$-Janus is a hybrid design of the above two. In order to probe into the stability of the studied Janus monolayers, the cohesive energy $E_{\text{coh}}$ is calculated by the following expression:

$$E_{\text{coh}} = [E_{\text{tot}} - (n_1 E_A + n_2 E_B + n_3 E_C + n_4 E_X + n_5 E_Y)]/(n_1 + n_2 + n_3 + n_4 + n_5)$$

where $E_{\text{tot}}$ is the total energy of the Janus monolayers. $E_A$, $E_B$, $E_C$, $E_X$, and $E_Y$ are the energies of isolated A, B, C, X, and Y atoms, respectively; $n_1$, $n_2$, $n_3$, $n_4$, and $n_5$ denote the number of A, B, C, X, and Y atoms in the unit cell, respectively. Our calculated results show that all Janus monolayers are energetically stable, as listed in Table SI. Moreover, the phonon spectra and AIMD simulations are calculated to verify their dynamical and thermal stabilities. For the sake of simplicity without sacrificing generality, we choose six materials, including the original systems Sb$_2$Si$_2$Te$_6$, Sb$_2$Ge$_2$Te$_6$, Bi$_2$Si$_2$Te$_6$, and one from each class of Janus structures (Sb$_2$Ge$_2$Se$_3$Te$_3$, Sb$_2$SiGeTe$_6$, Sb$_2$SiGeTe$_3$Se$_3$), as representative examples. All the three original systems and three representative Janus systems are dynamically stable, as shown in Fig. 1(c). One thing should be mentioned that a very small imaginary frequency (the largest imaginary frequency is −2.24 cm$^{-1}$) near the Γ-point is appeared, except the Sb$_2$Ge$_2$Te$_6$, Bi$_2$Si$_2$Te$_6$, and Sb$_2$Ge$_2$Se$_3$Te$_3$, resulting in relatively less dynamically stable of the remaining three monolayers. Imaginary frequency is a common problem in first-principles calculations of 2D materials, which could be easily affected by the supercell size, energy cutoff, or $k$-point sampling. The AIMD simulation at 300K shows that the free energy fluctuates in a narrow range, no chemical bond breaks and structural distortions occur throughout the process, which corroborates the thermal stability of the original and three representative Janus systems [Fig. 1(d)]. To verify the experimental realization of the Janus monolayers, we examine the total-energy differences between the final energy of Janus structures and the initial energy of A$_2$B$_2$Te$_6$, it is found that the transformation from A$_2$B$_2$Te$_6$ to the three types of Janus is mostly favorable, and show exothermic formation energy. However, the transformation from Bi$_2$Si$_2$Te$_6$ to Bi$_2$SiGeTe$_6$ and Bi$_2$Ge$_2$Se$_3$Te$_3$ is unfavorable, requiring energy to form Janus structures. Thus, substituting one of outer/ inner atomic layer or both inner and outer layers in the A$_2$B$_2$Te$_6$ monolayer with different elements is a simple way to achieve the C$_m$-Janus, I-Janus, and C$_p$-Janus structures. Actually, Jang *et al* successfully synthesized 3D Sb$_2$Si$_2$Te$_6$ and Bi$_2$Si$_2$Te$_6$ by a solid-state reaction method[37]. Due to system is a layered structure with A$_2$B$_2$Te$_6$ slabs stacked along the axis perpendicular to the A$_2$B$_2$Te$_6$ slab forming van der Waals bonds, which paves the way to obtaining a pure monolayer by advanced exfoliation, vapor deposition, or molecular-beam epitaxy techniques. Assuming the successful experimental synthesis of $A_2B_2Te_6$ monolayer, it is possible to get the Janus structures. For example, Lu *et al*. prepare $MoS_2$ monolayer by chemical vapor deposition, then strip off the top-layer sulfur atoms and replace them with hydrogen atoms using a remote hydrogen plasma, and then thermal selenization to replace H with Se to form a structurally stable Janus MoSSe monolayer[7]. Therefore, we look forward to the successful preparation of the $C_m$-Janus, I-Janus, and $C_p$-Janus structures.

**B. Electronic structure and tunability via external electric fields**

The band structures of 3 original monolayers and 30 Janus monolayers using PBE method with and without SOC are shown in Fig. S1. In the absence of SOC, for the 3 original systems, the conduction band minimum (CBM) is located at the Γ point, the valence band maximum (VBM) is located along the Γ - M direction (~1/2 Γ-M), possessing band gap of 1.04 eV, 0.67 eV, and 1.38 eV, respectively, which is almost consistent with previous results[26,27]. For $C_m$-Janus monolayers, except $Bi_2Ge_2S_3Te_3$ and $Bi_2Si_2S_3Te_3$ monolayers, the remaining 10 monolayers are indirect band gap semiconductors with a band gap range of 0.82-1.65 eV. Their CBM is located at the Γ point and VBM is along the Γ-M path. $Bi_2Ge_2S_3Te_3$ and $Bi_2Si_2S_3Te_3$ are direct band gap semiconductors, exhibiting band gap values of 1.28 eV and 1.76 eV, respectively. Both VBM and CBM of them are located at the Γ point. For I-Janus monolayers, the six monolayers are all indirect band gap semiconductors with a band gap range of 0.88-1.32 eV. Their CBM is located at the Γ point and the VBM is along the Γ-M path. For $C_p$-Janus monolayers, except for $Bi_2SiGeS_3Te_3$ and $Bi_2SiGeTe_3S_3$ monolayers, the remaining 10 monolayers are indirect band gap semiconductors with a band gap range of 1.03-1.49 eV. Their CBM is at the Γ point and the VBM is along the Γ-M path. $Bi_2SiGeS_3Te_3$ and $Bi_2SiGeTe_3S_3$ are direct band gap semiconductors, exhibiting band gap values of 1.51 eV and 1.59 eV, respectively. Both VBM and CBM of them are located at the Γ point. Fig. 2(a) presents the band structure of $C_m$-Janus $Sb_2Ge_2Se_3Te_3$ (SGST) monolayer without and with SOC effect included. When SOC is not considered, SGST monolayer is indirect band gap semiconductor, with corresponding band gap value of 0.82 eV. Its CBM is located at the Γ point and VBM is along the Γ-M path. When SOC is taken into account, the band gap decreases to 0.75 eV, and the large RSS appears at CBM due to the contributions of the $p_{x,y}$ and $p_z$ orbitals of chalcogen atoms (Se and Te) and the *s* orbital of Ge atom. To quantify the strength of the Rashba effect, the Rashba constant is defined as $\alpha_R = 2E_R/K_R$, where $E_R$ and $K_R$ are the Rashba energy and momentum offset, respectively[38,39]. Statistically, these 24 Rashba systems including SGST, cover a wide range of Rashba constant $\alpha_R$, varying from 0.06 to 0.86 eVÅ, as listed in Table SI. To further identify RSS, the spin texture is implemented with the relevant energy constants of +0.38 eV above the Fermi level, the opposite chiral circle of *in-plane* spin projection ($S_x$ and $S_y$) in Fig. 2(a) confirms the pure Rashba-type SOC. Considering the Rashba effect can be effectively modulated under $E_{ex}$, we apply $E_{ex}$ ($-0.3 \leq E_{ex} \leq 0.3$V/Å) to all three Janus types of Rashba systems[15,16,41,42]. It is found that the $\alpha_R$ increases or decreases linearly with the strength of the positive electric field, and correspondingly, the $\alpha_R$ under negative electric field maintains an opposite linear trend with the positive electric field range, which may be related to the direction of the $E_{in}$ in the monolayers[43] [Fig. 2(b)]. As shown in Fig. 2(c), the $A_2B_2X_3Y_3$ systems consist of two sets of $E_{in}$ from atom A to X (Y) and B to X (Y), which will enhance the strength of $E_{in}$ when the applied $E_{ex}$ is consistent with the $E_{in}$ direction, and weaken it otherwise. In addition to the SOC strength, the $E_{in}$ is generally the key factor affecting the $α_R$ of Janus Rashba systems and even its modulation trend under $E_{ex}$, which can directly reflected in aspects of work function, electrostatic potential, dipole moment, or inner charge transfer[24,44,45].

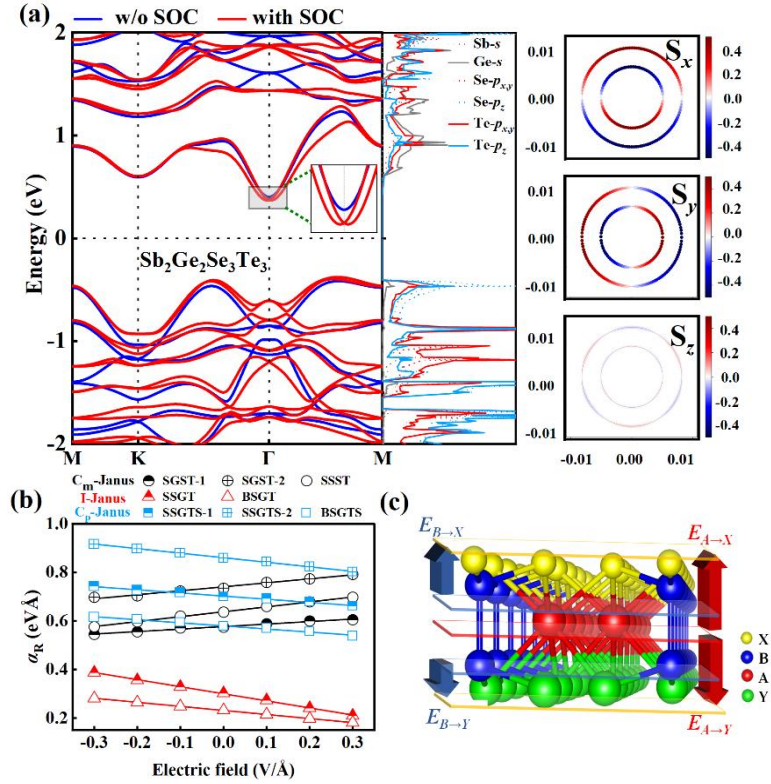

FIG. 2. (a) Band structure of SGST monolayer. Blue and red solid lines indicate methods of PBE and PBE+SOC, respectively. The Rashba spin splitting band is indicated in the magnified window on the right. The image on the right shows 2D contour plots of spin textures for SGST monolayer, where red and blue color represent spin up and spin down, respectively. (b) Electric field dependence of the Rashba constant $\alpha_R$ for the eight Janus monolayers: $Sb_2Ge_2S_3Te_3$ (SGST-1), $Sb_2Ge_2Se_3Te_3$ (SGST-2), $Sb_2Si_2Se_3Te_3$ (SSST), $Sb_2SiGeTe_6$ (SSGT), $Bi_2SiGeTe_6$ (BSGT), $Sb_2SiGeTe_3S_3$ (SSGTS-1), $Sb_2SiGeTe_3Se_3$ (SSGTS-2), $Bi_2SiGeTe_3Se_3$ (BSGTS). (c) Stereoscopic diagram of $A_2B_2X_3Y_3$ systems, where red and blue arrows represent the local electric field.

## C. Determinants of the Rashba constant: Physical mechanisms and their roles

Compared to the original structures, the Janus structures generally have different surface work functions and measurable effective potential energy differences due to the different types of elements on both sides, which can be estimated by the slope of potential energy differences ($k$) and work functions methods to determine the strength of its $E_{in}$[24]. For example, in the $C_m$-Janus systems we are studying, the correspondence between the slope $k$ calculated from the potential energy curves of these systems and their $α_R$ can be seen that in the $Sb_2B_2X_3Y_3$ and $Bi_2B_2X_3Y_3$ systems, the larger $α_R$ roughly corresponds to the larger $k$. It should be mentioned that compared to $Bi_2B_2X_3Y_3$, $Sb_2B_2X_3Y_3$ has a stronger Rashba effect due to its large $k$ [Fig. S3(b)]. Furthermore, we comprehensively calculate the work function, electrostatic potential and dipole moment of the three types of systems (Fig. S4 and Table SI). However, the three methods evaluating the strength of $E_{in}$ are not suitable for the I- and $C_p$-Janus systems, as they do not possess the typical characteristics of the $C_m$-Janus structures (limited to different elements on the outer sides). According to our previous research[16], the strength of the $E_{in}$ and $α_R$ can be effectively measured by charge transfer. As shown in Fig. 3($b_1$), the $A_2B_2X_6$ systems consist of upward local electric field $E_0$, $E_0'$ and downward local electric field $E_0$, $E_0'$. In the primitive systems $A_2B_2X_6$ (taking $Si_2Si_2Te_6$ as an example), the top ($-1.166\ e$, $e_0^- + e_{0'}^-$) and bottom ($-1.161\ e$, $e_0^- + e_{0'}^-$) layers receive approximately the same amount of charge from the middle layer, resulting in an $E_{in}$ of 0, because the system is symmetric and there is no difference in electronegativity between the atoms on both sides of the center [Fig. S3(c)]. The schematic diagram of the Janus structures (using $C_m$-Janus structures as an example) and the charge transfer between atoms in adjacent layers are shown in Fig. 3($b_2$). For Janus structures ($C_m$-Janus $Sb_2Ge_2Se_3Te_3$, I-Janus $Si_2SiGeTe_6$, and $C_p$-Janus $Sb_2SiGeTe_3Se_3$), the top layer ($-1.603\ e$, $-1.162\ e$, and $-1.127\ e$, $e_1^- + e_3^-$) and the bottom layer ($-0.857\ e$, $-0.869\ e$, and $-1.604\ e$, $e_2^- + e_4^-$) gain different amounts of charge from the middle layer due to the different electronegativity, and the charge transfer difference ($Q_d$) between the two ends ($0.746\ e$, $0.293\ e$, and $0.477\ e$, $Q_d = |e_1^- + e_3^- - e_2^- - e_4^-|$) creates a potential gradient in the base plane, resulting in an $E_{in}$ between the top and bottom layers. In addition, the magnitude of the electronegativity difference between elements determines the total charge transfer ($Q_t$), for example, the greater the electronegativity difference between adjacent elements, the greater the change in $Q_t$, and $Q_t$, $Q_d$ jointly affect the $E_{in}$. From the most fundamental perspective, the SOC strength of the system is also a key factor that determines the $\alpha_R$[46]. The SOC can be reflected by atomic number, which has been verified by the fact that the SOC strength of TMD monolayers increases with the increases of chalcogen atomic number[47]. Taking the SSGTS-2 monolayer in Fig. S3(c) as an example, although $Q_d$ is not large at 0.477 $e$, it has a high atomic number in the $C_p$-Janus structures $A_2BCX_3Y_3$ (Excluding the common element Sb, the atomic number sum is 304), leading to the monolayer still possessing the significant Rashba effect. Due to the special construction of replacing multiple atoms in the design of such Janus structures, resulting in an obvious change in the total atomic number, unlike the conventional a single atom substitution, which leads to an obvious influence of the atomic number on the Rashba effect. In summary, due to the large SOC and the existence of $E_{in}$, the atomic number and charge transfer are key factors influencing the strength of the Rashba effect [Fig. 3(a)].

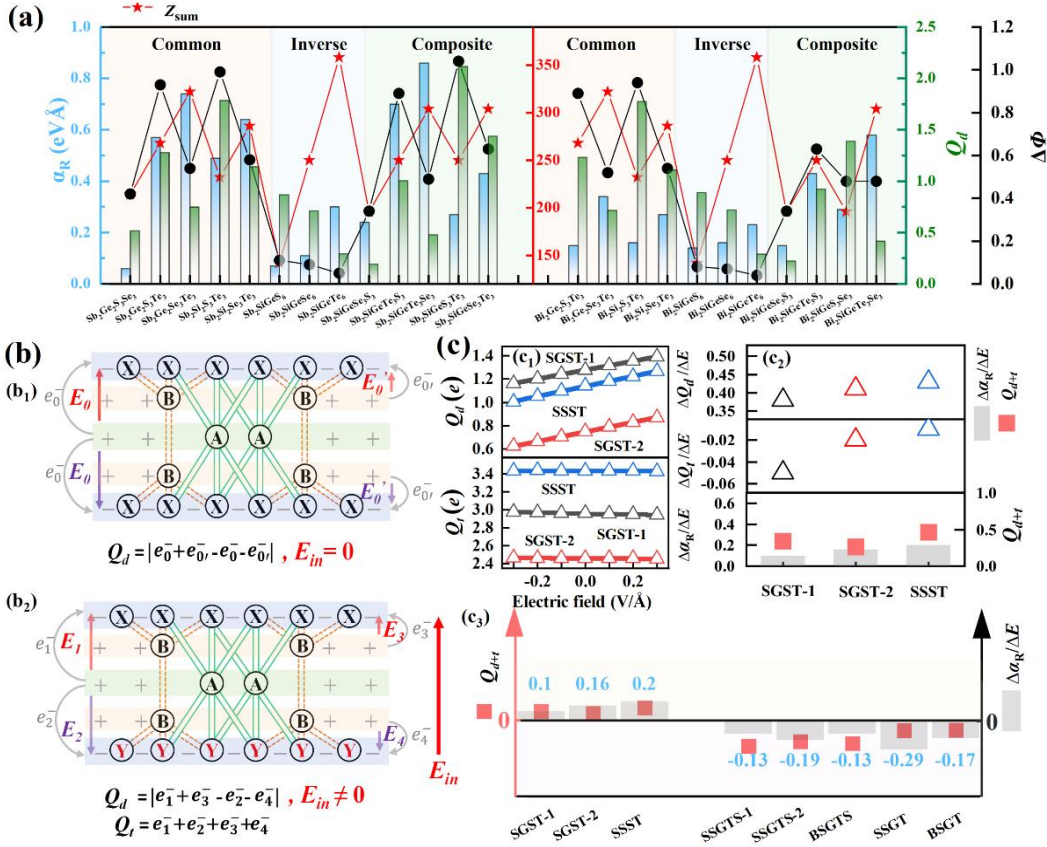

FIG. 3. (a) The Rashba constant $\alpha_R$, sum of atomic numbers $Z_{sum}$, charge transfer difference $Q_d$, and work function difference $\Delta\Phi$ of the Janus structures. (b) Schematic diagram of charge transfer of ($b_1$) $A_2B_2X_6$ and ($b_2$) $A_2B_2X_3Y_3$. The $A_2B_2X_6$ systems can be divided into individual components $A_2X_6$ and $B_2X_6$, represented by green solid lines and yellow dashed lines, respectively. $e_0^-, e_{0'}^-, e_1^-, e_2^-, e_3^-, e_4^-$ represent the flow of electrons between different layers. $E_0$, $E_0'$, $E_1$, $E_2$, $E_3$, $E_4$ represent the localized electric field. (c) ($c_1$) Electric field dependence of $Q_d$ and $Q_t$ for the three $C_m$-Janus monolayers: SGST-1, SGST-2, SSST. ($c_2$) The electric field response rate $\Delta\alpha_R/\Delta E$ and $Q_{d+t}$ of $C_m$-Janus structures. ($c_3$) $Q_{d+t}$ and slope $\Delta\alpha_R/\Delta E$ of 8 monolayers under $E_{ex}$, where the value represents the electric field response rate.

Due to the fact that charge transfer in the Janus structures is one of the reasons for the formation of the Rashba effect, we speculate that the change in charge transfer under $E_{ex}$ may be related to the strength of RSS change. The Fig. 3(c₁) illustrates the linear variation of $Q_d$ and $Q_t$ under $E_{ex}$, we define their electric field responses as $k_1 = \frac{\Delta Q_d}{\Delta E}$ and $k_2 = \frac{\Delta Q_t}{\Delta E}$, respectively. Considering that $Q_d$ can reflect the strength of $E_{in}$ by generating a potential gradient in the basal plane at both ends of monolayers, $Q_t$ reflects the strength of the electronegativity difference between atoms by the total amount of charge transfer. To explain the change in RSS of Janus structures under the action of $E_{ex}$, we continue to use the mathematical expression $Q_d \times k_1 + Q_t \times k_2$ based on $Q_d$ and $Q_t$, which is abbreviated as $Q_{d+t}$[16].

$$Q_{d+t} \propto \frac{\Delta \alpha_R}{\Delta E}$$

For the charge transfer of $C_m$-Janus structures under $E_{ex}$, we obtain the relationship between $\frac{\Delta \alpha_R}{\Delta E}$ and $Q_{d+t}$ and extend it to I- and $C_p$-Janus structures [Fig. 3(c₂)]. As a result, the numerical values of $Q_{d+t}$ are 0.34, 0.26 and 0.46 for SGST-1, SGST-2 and SSST monolayers; The $C_p$-Janus structures SSGTS-1, SSGTS-2, BSGTS, and I-Janus structures SSGT, BSGT have values of -1.03, -0.86, -0.93, -0.42 and -0.39, respectively. As shown in Fig. 3(c₃), when $Q_{d+t}$ is greater than 0, the RSS enhances (weakens) monotonically with the increase (decrease) of $E_{ex}$; When $Q_{d+t}$ is less than 0, the RSS weakens (enhances) monotonically with the increase (decrease) of $E_{ex}$. Therefore, when combining the $Q_d$ with another $Q_t$ related to element electronegativity, the trend of $\alpha_R$ increasing or decreasing with the strength of positive electric field can be further approximately confirmed based on the slope of charge transfer change under $E_{ex}$.

**D. Charge transfer-mediated regulation of the Rashba effect**

Inspired by the quantitative relationship between charge transfer and the strength of $E_{in}$ and even the $\alpha_R$, there are two thought-provoking problems as follows: Could the $E_{in}$ be designed from the perspective of charge transfer to introduce Rashba effect in symmetrical structure? And could we regulate the $E_{in}$ via the further design of geometric structures based on Janus structures? As discussed below, the straightforward approach is introducing a single adatom onto the surface of 2D monolayer. Next, we examine the electronic structure and Rashba effect of a single atom adsorption onto pure $Sb_2Ge_2Te_6$ (SGT) and Janus SGST systems.

We choose the S atom with strong electronegativity as the adatom and consider five possible adsorption sites as shown in Fig. 4(a$_1$), named B, H$_1$, H$_2$, T$_{Ge}$, and T$_{Se/Te}$, respectively. The adsorption energy ($E_{ads}$) can be calculated as follows:

$$E_{\text{ads}} = E_{S-substrate} - (E_{substrate} + E_S)$$

where $E_{S\text{-}substrate}$, $E_{substrate}$, and $E_S$ are the total energies of the S-SGT(S-SGST) system, SGT(SGST) monolayer, and S atom, respectively. As shown in Fig. 4(a$_2$), in the S-SGT system, the T$_{Te}$ site has the lowest adsorption energy of -1.92 eV; In the S-SGST system, when S adsorbs onto the Se (Te) surface, the T$_{Se}$ (T$_{Te}$) site has a minimum adsorption energy of -1.73 eV (-1.80 eV). Consequently, only the Rashba effect at the optimal T$_{Se/Te}$ position are investigated in the following discussions. When an S atom is adsorbed on pure SGT monolayer, the structural symmetry of the monolayer is broken, and significant RSS occurs at the CBM ($\alpha_R^K$ = 0.89 eVÅ, $\alpha_R^M$ = 0.51 eVÅ). Interestingly, we find that anisotropic RSS also occurs at the VBM ($\alpha_R^K$ = 0.96 eVÅ, $\alpha_R^M$ = 0.61 eVÅ), and it is very rare to have RSS at both the CBM and VBM [Fig. 4(b$_1$)]. The anisotropy of the Rashba effect caused by the addition of adsorbed atoms has also been reported in previous works[48]. The charge density difference of the S-SGT system is computed, and the Bader charge is further used to determine that the S atom obtains 0.541 $e$ from the Te layer, resulting in a non-zero local electric field ($E_{Te\text{-}S}$) in the system [Fig. 4(c$_1$)]. When the S atom is adsorbed on the SGST (Se-surface), the $\alpha_R$ at the CBM is 1.19/0.85 eVÅ, which is higher than the 0.74 eVÅ of the SGST monolayer [Fig. 4(b$_2$)]. Bader charge shows that the S atom obtains 0.337 $e$ from the Se layer, which leads to the redistribution of charge and generates a local electric field in the direction of the Se layer towards the S atom. This is consistent with the direction of the $E_{in}$ in the SGST monolayer, resulting in the enhancement of the $E_{in}$ of the system and increasing the $\alpha_R$ [Fig. 4(c$_2$)]. When adsorbed on the Te surface of SGST, the S atom obtains 0.542 $e$ from the Te layer, and the direction of the local electric field from the Te layer to the S atom is opposite to the $E_{in}$ of the SGST monolayer [Fig. 4(c$_3$)], which leads to the weakening of the $E_{in}$ in the system, and the $\alpha_R$ decreases to 0.40/0.34 eVÅ at CBM [Fig. 4(b$_3$)]. Notable, the S-SGST system exhibits a transition from indirect band gap to direct band gap compared to the SGST monolayer, and also shows RSS at the VBM with $\alpha_R^K$ = 0.64 eVÅ, $\alpha_R^M$ = 0.38 eVÅ for the Se-surface, and $\alpha_R^K$ = 0.59 eVÅ, $\alpha_R^M$ = 0.39 eVÅ for the Te-surface [Fig. 4(b$_2$) and 4(b$_3$)]. The reason for such Rashba splitting in this VBM can be ascribed to the presence of S atomic orbital near the Fermi level as depicted by the projected DOS plots for three adsorption systems (Fig. S5). The above results indicate that the reconfiguration of charge transfer through the adsorption of atom can be an effective method to introduce and regulate the Rashba effect. Furthermore, we further investigate the adsorption

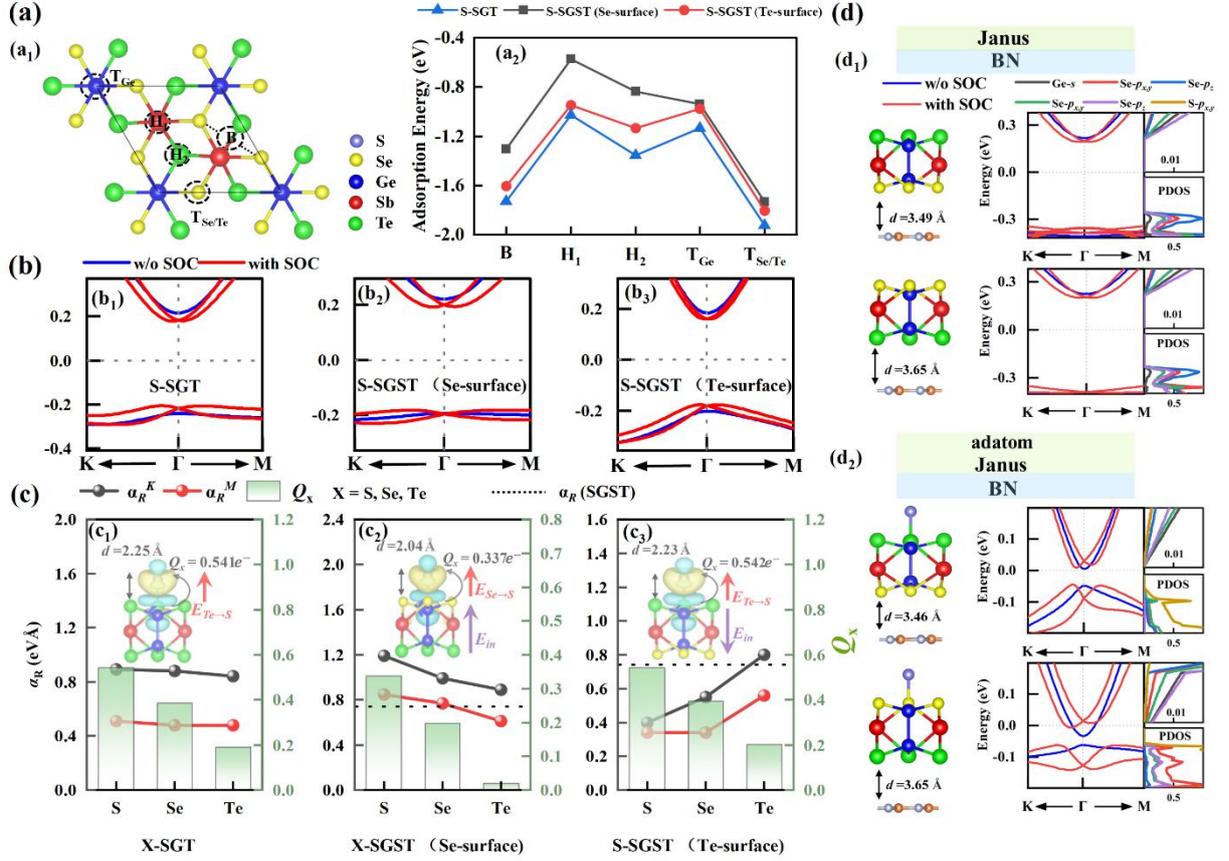

FIG. 4. (a$_1$) Adsorbable position of pure SGT and Janus SGST monolayer. (a$_2$) The adsorption energies ($E_{ads}$) of a single S atom adsorbed on pure SGT and Janus SGST monolayer for five typical adsorption positions. (b) Enlarged view of RSS bands of (b$_1$) S-SGT, (b$_2$) S-SGST (Se-surface) and (b$_3$) S-SGST (Te-surface) systems near Fermi level. (c) Charge transfer $Q_x$ and the strength of Rashba effect of (c$_1$) X adsorbed on pure SGT, (c$_2$) X adsorbed on SGST (Se-surface) and (c$_3$) X adsorbed on SGST (Te-surface). The middle part is the charge density difference of the S-SGT/SGST system, where the red arrow represents the local electric field of the monolayer and the adsorbed atom, and the purple arrow represents the $E_{in}$ of the monolayer. The yellow and blue regions represent the electron accumulation and depletion, respectively. $d$ represents the vertical distance between the S atom and the adjacent Te/Se layer. (d) The structure, energy band, and pdos near the Fermi level of (d$_1$) the BN/SGST heterostructures and (d$_2$) the BN/S-SGST

behavior of X (X = S, Se, Te) elements, which belong to the same main group as the S element, on the pure SGT and Janus SGST systems (Fig. S6). It is found that the surface electric field strength, primarily driven by charge transfer, decreased with the decrease in the electronegativity of the adsorbed elements. This, in conjunction with the intrinsic electric field within the pristine 2D system, collectively influenced the magnitude of the $α_R$ at the CBM [Fig. 4(c)].

At the end, it is necessary to explore the feasibility of 2D materials growth on substrates and the influence of substrates on electronic properties. To minimize lattice mismatch as much as possible, we establish heterostructures using a 3 × 3 × 1 supercell of BN and SGST (S-SGST) monolayer. The lattice constant of SGST is 7.06 Å, the lattice constant of BN is 2.50 Å, and the lattice mismatch calculated by $f = (\alpha_s - \alpha_g)/\alpha_s$ is around 6%, where $\alpha_s$ and $\alpha_g$ are the original lattice constants of the seed and overgrowth material, respectively[49]. For the four heterostructures of BN/SGST (Se and Te-surface) and BN/S-SGST (Se and Te-surface), we validate their stability through AIMD simulations with small-scale energy fluctuations [Fig. S7(c)]. When SOC is taken into account, the RSS at CBM of BN/SGST increases to 0.78 eVÅ (Se-surface) and 0.81 eVÅ (Te-surface), respectively, which is only slightly above the value of 2D SGST monolayer (0.74 eVÅ) [Fig. 4($d_1$)].

TABLE I. For S-SGT/SGST adsorption and BN/S-SGST heterostructure systems: band gap without and with SOC ($E_{g\text{-PBE}}$ and $E_{g\text{-PBE+SOC}}$), the Rashba constant $\alpha_R^K/\alpha_R^M$ along the Γ-K/M direction at the conduction band Γ point and the Rashba constant $\alpha_R^K/\alpha_R^M$ along the Γ-K/M direction at the valence band Γ point.

| Systems | Interface | $E_{g\text{-PBE}/\text{PBE+SOC}}$ (eV) | CBM $\alpha_R^K/\alpha_R^M$ (eVÅ) | VBM $\alpha_R^K/\alpha_R^M$ (eVÅ) |
|---|---|---|---|---|
| S-SGT | Te-surface | 0.46/0.38 | 0.89/0.51 | 0.96/0.61 |
| S-SGST | Se-surface | 0.37/0.34 | 1.19/0.85 | 0.64/0.38 |
|  | Te-surface | 0.39/0.34 | 0.40/0.34 | 0.59/0.39 |
| BN/S-SGST | Se-surface | 0.05/0.05 | 2.17/1.01 | 2.88/1.74 |
|  | Te-surface | 0.03/0.05 | 1.46/0.77 | 2.05/1.43 |

For BN/S-SGST heterostructures, the strength of the anisotropic RSS along the Γ-K and Γ-M directions at CBM is $\alpha_R^K$ = 2.17 eVÅ and $\alpha_R^M$= 1.01 eVÅ for the Se-surface and $\alpha_R^K$ = 1.46 eVÅ and $\alpha_R^M$ = 0.77 eVÅ for the Te-surface [Fig. 4($d_2$) and Table I]. In addition, the anisotropic RSS along the Γ-K and Γ-M directions at VBM is significantly enhanced, where $\alpha_R^K$ = 2.88 eVÅ and $\alpha_R^M$ = 1.74 eVÅ for the Se-surface and $\alpha_R^K$ = 2.05 eVÅ and $\alpha_R^M$ = 1.43 eVÅ for the Te-surface (Table I). The VBM is mainly contributed by the $p_{x,y}$ orbital of the S atom and the $p_{x,y}$ and $p_z$ orbitals of chalcogen atoms (Se and Te) for BN/S-SGST, resulting in the large Rashba effect. Such 2D giant Rashba effect can pave the way to use the BN/S-SGST heterostructure systems in the development of highly efficient SFETs, along with the spintronics industry.

## IV. Conclusion

In summary, we theoretically design and predict a series of Janus monolayers with RSS, and demonstrate that the magnitude of Rashba constant $\alpha_R$ and changes under $E_{ex}$ are closely related to charge transfer. More notably, based on the mechanism of charge transfer, we successfully introduce anisotropic RSS into symmetric structure SGT through the most direct method of introducing a single adatom onto the surface of 2D monolayer, and modulate $α_R$ by selecting chalcogen elements with different electronegativities for adsorption in the Janus SGST. Lastly, we examine the growth feasibility and electronic structures of the Janus system and Janus-adsorbed systems on a 2D BN substrate. Our work not only presents a series of 2D Rashba semiconductors with potential in SFETs, but also particularly proposes a simple strategy for controllably modulating the RSS via charge transfer.

## Author contributions

Z.Q. conceived and designed the research. H.W. carried out the calculations and analyzed the calculated results. G.Q. participated in the discussion and provided insightful suggestions. All the authors contributed to the final revision of this article.

## Conflicts of interest

The authors declare no competing interests.

## Acknowledgements

Z.Q. is supported by the National Natural Science Foundation of China (Grant No.12274374). G. Q. is supported by the National Natural Science Foundation of China (Grant No. 52006057), the Fundamental Research Funds for the Central Universities (Grant No. 531119200237 and 541109010001). The numerical calculations in this work are supported by National Supercomputing Center in Zhengzhou.

## References

1. Manchon, A., Koo, H. C., Nitta, J., Frolov, S. M. & Duine, R. A. New perspectives for Rashba spin–orbit coupling. *Nature Mater* **14**, 871–882 (2015).
2. Meier, L. *et al.* Measurement of Rashba and Dresselhaus spin–orbit magnetic fields. *Nature Phys* **3**, 650–654 (2007).
3. Awschalom, D. & Samarth, N. Spintronics without magnetism. *Physics* **2**, 50 (2009).
4. Dyrdał, A. & Barnaś, J. Current-induced spin polarization and spin-orbit torque in graphene. *Phys. Rev. B* **92**, 165404 (2015).
5. Ganichev, S. D. *et al.* Spin-galvanic effect due to optical spin orientation in $n$-type GaAs quantum well structures. *Phys. Rev. B* **68**, 081302 (2003).

6. Ghobadi, N. & Touski, S. B. The Electrical and Spin Properties of Monolayer and Bilayer Janus HfSSe under Vertical Electrical Field. *J. Phys.: Condens. Matter* **33**, 085502 (2021).

7. Lu, A.-Y. *et al.* Janus monolayers of transition metal dichalcogenides. *Nature Nanotech* **12**, 744–749 (2017).

8. Montes-García, V. & Samorì, P. Janus 2D materials *via* asymmetric molecular functionalization. *Chem. Sci.* **13**, 315–328 (2022).

9. Zhang, L. *et al.* Recent advances in emerging Janus two-dimensional materials: from fundamental physics to device applications. *J. Mater. Chem. A* **8**, 8813–8830 (2020).

10. Wu, K. *et al.* Two-Dimensional Giant Tunable Rashba Semiconductors with Two-Atom-Thick Buckled Honeycomb Structure. *Nano Lett.* **21**, 740–746 (2021).

11. Chen, J., Wu, K., Ma, H., Hu, W. & Yang, J. Tunable Rashba spin splitting in Janus transition-metal dichalcogenide monolayers *via* charge doping. *RSC Adv.* **10**, 6388–6394 (2020).

12. Zhu, L., Zhang, T., Chen, G. & Chen, H. Huge Rashba-type spin–orbit coupling in binary hexagonal PX nanosheets (X = As, Sb, and Bi). *Phys. Chem. Chem. Phys.* **20**, 30133–30139 (2018).

13. Hu, T. *et al.* Intrinsic and anisotropic Rashba spin splitting in Janus transition-metal dichalcogenide monolayers. *Phys. Rev. B* **97**, 235404 (2018).

14. Liu, C. *et al.* Manipulation of the Rashba effect in layered tellurides MTe (M = Ge, Sn, Pb). *J. Mater. Chem. C* **8**, 5143–5149 (2020).

15. Liu, M.-Y., Gong, L., He, Y. & Cao, C. Tuning Rashba effect, band inversion, and spin-charge conversion of Janus X Sn 2 Y monolayers via an external field. *Phys. Rev. B* **103**, 075421 (2021).

16. Tian, Q. *et al.* Inverse Janus design of two-dimensional Rashba semiconductors. *Phys. Rev. B* **108**, 115130 (2023).

17. Ghobadi, N., Gholami Rudi, S. & Soleimani-Amiri, S. Electronic, spintronic, and piezoelectric properties of new Janus Zn A X Y ( A = Si , Ge , Sn , and X , Y = S , Se , Te ) monolayers. *Phys. Rev. B* **107**, 075443 (2023).

18. Kong, W. *et al.* Tunable Rashba spin-orbit coupling and its interplay with multiorbital effect and magnetic ordering at oxide interfaces. *Phys. Rev. B* **104**, 155152 (2021).

19. Guo, S.-D. *et al.* Switching Rashba spin-splitting by reversing electric-field direction. *Phys. Rev. Materials* **7**, 044604 (2023).

20. Mohanta, M. K., Is, F., Kishore, A. & De Sarkar, A. Spin-Current Modulation in Hexagonal Buckled ZnTe and CdTe Monolayers for Self-Powered Flexible-Piezo-Spintronic Devices. *ACS Appl. Mater. Interfaces* **13**, 40872–40879 (2021).

21. Is, F., Mohanta, M. K. & Sarkar, A. D. Insights into selected 2D piezo Rashba semiconductors for self-powered flexible piezo spintronics: material to contact properties. *J. Phys.: Condens. Matter* **35**, 253001 (2023).

22. Qin, Z., Qin, G., Shao, B. & Zuo, X. Rashba spin splitting and perpendicular magnetic anisotropy of Gd-adsorbed zigzag graphene nanoribbon modulated by edge states under external electric fields. *Phys. Rev. B* **101**, 014451 (2020).

23. Chen, S. *et al.* Large tunable Rashba spin splitting and piezoelectric response in Janus chromium dichalcogenide monolayers. *Phys. Rev. B* **106**, 115307 (2022).

24. Wang, D. *et al.* Dipole control of Rashba spin splitting in a type-II Sb/InSe van der Waals heterostructure. *J. Phys.: Condens. Matter* **33**, 045501 (2021).

25. Li, T., Liu, J., Sun, Q., Kawazoe, Y. & Jena, P. A record high average ZT over a wide temperature range in a Single-layer Sb2Si2Te6. *Applied Surface Science* **567**, 150873 (2021).

26. Shi, W., Ge, N., Wang, X. & Wang, Z. High Thermoelectric Performance of $Sb_2Si_2Te_6$ Monolayers. *J. Phys. Chem. C* **125**, 16413–16419 (2021).

27. Zhang, T., Ning, S., Zhang, Z., Qi, N. & Chen, Z. Dimensionality reduction induced synergetic optimization of the thermoelectric properties in $Bi_2Si_2X_6$ (X = Se, Te) monolayers. *Phys. Chem. Chem. Phys.* **25**, 25029–25037 (2023).

28. Kresse, G. & Furthmüller, J. Efficient iterative schemes for *ab initio* total-energy calculations using a plane-wave basis set. *Phys. Rev. B* **54**, 11169–11186 (1996).

29. Blöchl, P. E. Projector augmented-wave method. *Phys. Rev. B* **50**, 17953–17979 (1994).

30. Perdew, J. P., Burke, K. & Ernzerhof, M. Generalized Gradient Approximation Made Simple. *Phys. Rev. Lett.* **77**, 3865–3868 (1996).

31. Monkhorst, H. J. & Pack, J. D. Special points for Brillouin-zone integrations. *Phys. Rev. B* **13**, 5188–5192 (1976).

32. Togo, A. & Tanaka, I. First principles phonon calculations in materials science. *Scripta Materialia* **108**, 1–5 (2015).

33. Tang, W., Sanville, E. & Henkelman, G. A grid-based Bader analysis algorithm without lattice bias. *J. Phys.: Condens. Matter* **21**, 084204 (2009).

34. Momma, K. & Izumi, F. *VESTA 3* for three-dimensional visualization of crystal, volumetric and morphology data. *J Appl Crystallogr* **44**, 1272–1276 (2011).

35. Babaee Touski, S. & Ghobadi, N. Structural, electrical, and Rashba properties of monolayer Janus $Si_2XY$ (X, Y =P, As, Sb, and Bi). *Phys. Rev. B* **103**, 165404 (2021).

36. Petrić, M. M. *et al.* Raman spectrum of Janus transition metal dichalcogenide monolayers WSSe and MoSSe. *Phys. Rev. B* **103**, 035414 (2021).

37. Jang, H. *et al.* The Comparative Study of Thermoelectric Properties in Sb2Si2Te6 and Bi2Si2Te6.

38. Rezavand, A. & Ghobadi, N. Stacking-dependent Rashba spin-splitting in Janus bilayer transition metal dichalcogenides: The role of in-plane strain and out-of-plane electric field. *Physica E: Low-dimensional Systems and Nanostructures* **132**, 114768 (2021).

39. Yao, Q.-F. *et al.* Manipulation of the large Rashba spin splitting in polar two-dimensional transition-metal dichalcogenides. *Phys. Rev. B* **95**, 165401 (2017).

40. Rezavand, A. & Ghobadi, N. First-principle study on quintuple-atomic-layer Janus $MTeSiX_2$ (M= Mo, W; X=N, P, As) monolayers with intrinsic Rashba spin-splitting and Mexican hat dispersion. *Materials Science in Semiconductor Processing* **152**, 107061 (2022).

41. Caviglia, A. D. *et al.* Tunable Rashba Spin-Orbit Interaction at Oxide Interfaces. *Phys. Rev. Lett.* **104**, 126803 (2010).

42. Domaretskiy, D. *et al.* Quenching the bandgap of two-dimensional semiconductors with a perpendicular electric field. *Nat. Nanotechnol.* **17**, 1078–1083 (2022).

43. Ju, W. *et al.* Remarkable Rashba spin splitting induced by an asymmetrical internal electric field in polar III–VI chalcogenides. *Phys. Chem. Chem. Phys.* **22**, 9148–9156 (2020).

44. Yu, S.-B., Zhou, M., Zhang, D. & Chang, K. Spin Hall effect in the monolayer Janus compound MoSSe enhanced by Rashba spin-orbit coupling. *Phys. Rev. B* **104**, 075435 (2021).

45. Chen, J., Wu, K., Ma, H., Hu, W. & Yang, J. Tunable Rashba spin splitting in Janus transition-metal dichalcogenide monolayers *via* charge doping. *RSC Adv.* **10**, 6388–6394 (2020).

46. Yuan, J. *et al.* One-dimensional thermoelectrics induced by Rashba spin-orbit coupling in two-dimensional BiSb monolayer. *Nano Energy* **52**, 163–170 (2018).

47. Zibouche, N., Kuc, A., Musfeldt, J. & Heine, T. Transition-metal dichalcogenides for spintronic

applications. *Annalen der Physik* **526**, 395–401 (2014).

48. Absor, Moh. A. U., Ishii, F., Kotaka, H. & Saito, M. Spin-split bands of metallic hydrogenated ZnO (10$\bar{1}$) surface: First-principles study. *AIP Advances* **6**, 025309 (2016).

49. Huda, M. N. & Kleinman, L. h - BN monolayer adsorption on the Ni ( 111 ) surface: A density functional study. *Phys. Rev. B* **74**, 075418 (2006).

*Supplemental Material*

# Rashba effect modulation in two-dimensional $A_2B_2Te_6$ (A = Sb, Bi; B = Si, Ge) materials via charge transfer

Haipeng Wu[1], Qikun Tian[2], Jinghui Wei[1], Ziyu Xing[1], Guangzhao Qin[2], Zhenzhen Qin[1*]

[1]*Key Laboratory of Materials Physics, Ministry of Education, School of Physics, Zhengzhou University, Zhengzhou 450001, P. R. China*

[2]*State Key Laboratory of Advanced Design and Manufacturing Technology for Vehicle, College of Mechanical and Vehicle Engineering, Hunan University, Changsha 410082, P. R. China*

* Corresponding author: qzz@zzu.edu.cn

## A. Structural parameters of original and Janus monolayers

Table SI shows the optimized lattice parameters of original and Janus structures, the lattice parameters of the $C_m$-Janus $A_2B_2X_3Y_3$ monolayers range from 6.59 to 7.15 Å, where the $Bi_2Ge_2Se_3Te_3$ monolayer having the maximum lattice parameter of 7.15 Å; The lattice parameters of the I-Janus $A_2BCX_6$ monolayers range from 6.49 to 7.36 Å, where the $Bi_2SiGeTe_6$ monolayer having the maximum lattice parameter of 7.36 Å; The lattice parameters of the $C_p$-Janus $A_2BCX_3Y_3$ monolayers range from 6.63 to 7.13 Å, where the $Bi_2SiGeSe_3Te_3$ monolayer having the maximum lattice parameter of 7.13 Å. The variation of lattice parameters can be attributed to the size of atomic radius, therefore the structure with the largest atomic radius in each class of Janus has the largest lattice parameter.

TABLE SI. For original and Janus monolayers: lattice constant ($a = b$), band gap without and with SOC ($E_{\text{g-PBE}}$ and $E_{\text{g-PBE+SOC}}$), cohesive energy ($E_{\text{coh}}$), position and constant $\alpha_R$ of Rashba effect ($C_1$ and $V_1$ represent the conduction band and valence band closest to the Fermi level, respectively.), dipole moment ($P$).

| Classes | Systems | $a=b$ (Å) | $E_{\text{g-PBE}}$ (eV) | $E_{\text{g-PBE+SOC}}$ (eV) | $E_{\text{coh}}$ (eV/atom) | Position ($\Gamma$) | $\alpha_R$ (eVÅ) | $P$ (e×Å) |
|---|---|---|---|---|---|---|---|---|
| $A_2B_2Te_6$ | $Sb_2Si_2Te_6$ | 7.23 | 1.04(i) | 0.90 | -3.18 | / | 0 | 0 |
| | $Sb_2Ge_2Te_6$ | 7.29 | 0.67(i) | 0.58 | -2.99 | / | 0 | 0 |
| | $Bi_2Si_2Te_6$ | 7.33 | 1.38(i) | 0.83 | -3.22 | / | 0 | 0 |
| $A_2B_2X_3Y_3$ | $Sb_2Ge_2S_3Se_3$ | 6.68 | 0.90(i) | 0.84 | -3.36 | $C_1/V_1$ | 0.06/0.41 | 0.089 |
| | $Sb_2Ge_2S_3Te_3$ | 6.94 | 1.12(i) | 0.99 | -3.22 | $C_1$ | 0.57 | 0.213 |
| | $Sb_2Ge_2Se_3Te_3$ | 7.06 | 0.82(i) | 0.75 | -3.12 | $C_1$ | 0.74 | 0.129 |
| | $Sb_2Si_2S_3Se_3$ | 6.59 | 1.33(i) | 1.23 | -3.65 | $V_1$ | 0.34 | 0.093 |
| | $Sb_2Si_2S_3Te_3$ | 6.87 | 1.50(i) | 1.34 | -3.47 | $C_1$ | 0.49 | 0.223 |
| | $Sb_2Si_2Se_3Te_3$ | 6.99 | 1.20(i) | 1.08 | -3.34 | $C_1$ | 0.64 | 0.134 |
| | $Bi_2Ge_2S_3Se_3$ | 6.78 | 1.24(i) | 0.96 | -3.42 | $V_1$ | 0.47 | 0.089 |
| | $Bi_2Ge_2S_3Te_3$ | 7.04 | 1.28(d) | 0.73 | -3.27 | $C_1$ | 0.15 | 0.211 |
| | $Bi_2Ge_2Se_3Te_3$ | 7.15 | 1.12(i) | 0.74 | -3.16 | $C_1$ | 0.34 | 0.127 |
| | $Bi_2Si_2S_3Se_3$ | 6.70 | 1.65(i) | 1.22 | -3.70 | $V_1$ | 0.41 | 0.091 |
| | $Bi_2Si_2S_3Te_3$ | 6.97 | 1.76(d) | 1.02 | -3.52 | $C_1$ | 0.16 | 0.218 |
| | $Bi_2Si_2Se_3Te_3$ | 7.09 | 1.51(i) | 1.02 | -3.38 | $C_1$ | 0.27 | 0.131 |
| $A_2BCX_6$ | $Sb_2SiGeS_6$ | 6.49 | 1.32(i) | 1.27 | -3.63 | $C_1$ | 0.07 | 0.022 |
| | $Sb_2SiGeSe_6$ | 6.78 | 0.94(i) | 0.88 | -3.38 | $C_1/V_1$ | 0.11/0.17 | 0.019 |
| | $Sb_2SiGeTe_6$ | 7.26 | 0.88(i) | 0.75 | -3.08 | $C_1$ | 0.30 | 0.013 |
| | $Bi_2SiGeS_6$ | 6.60 | 1.63(i) | 1.28 | -3.69 | $C_1$ | 0.14 | 0.016 |
| | $Bi_2SiGeSe_6$ | 6.87 | 1.28(i) | 0.97 | -3.43 | $C_1/V_1$ | 0.16/0.30 | 0.014 |
| | $Bi_2SiGeTe_6$ | 7.36 | 1.21(i) | 0.73 | -3.12 | $C_1$ | 0.23 | 0.011 |
| $A_2BCX_3Y_3$ | $Sb_2SiGeSe_3S_3$ | 6.63 | 1.15(i) | 1.07 | -3.49 | $C_1/V_1$ | 0.24/0.25 | 0.070 |
| | $Sb_2SiGeTe_3S_3$ | 6.90 | 1.33(i) | 1.19 | -3.31 | $C_1$ | 0.70 | 0.202 |
| | $Sb_2SiGeS_3Se_3$ | 6.64 | 1.14(i) | 1.06 | -3.52 | $V_1$ | 0.51 | 0.111 |
| | $Sb_2SiGeTe_3Se_3$ | 7.02 | 1.03(i) | 0.93 | -3.21 | $C_1$ | 0.86 | 0.116 |
| | $Sb_2SiGeS_3Te_3$ | 6.91 | 1.33(i) | 1.18 | -3.37 | $C_1$ | 0.27 | 0.237 |
| | $Sb_2SiGeSe_3Te_3$ | 7.03 | 1.04(i) | 0.93 | -3.24 | $C_1$ | 0.43 | 0.148 |
| | $Bi_2SiGeSe_3S_3$ | 6.74 | 1.49(i) | 1.13 | -3.54 | $C_1/V_1$ | 0.15/0.28 | 0.075 |
| | $Bi_2SiGeTe_3S_3$ | 7.00 | 1.59(d) | 0.93 | -3.36 | $C_1$ | 0.43 | 0.202 |
| | $Bi_2SiGeS_3Se_3$ | 6.75 | 1.46(i) | 1.11 | -3.58 | $C_1/V_1$ | 0.29/0.58 | 0.105 |

| | | | | | | | |
|---|---|---|---|---|---|---|---|
| Bi$_2$SiGeTe$_3$Se$_3$ | 7.12 | 1.34(i) | 0.93 | -3.26 | C$_1$ | 0.58 | 0.116 |
| Bi$_2$SiGeS$_3$Te$_3$ | 7.01 | 1.51(d) | 0.85 | -3.42 | / | 0 | 0.227 |
| Bi$_2$SiGeSe$_3$Te$_3$ | 7.13 | 1.34(i) | 0.87 | -3.29 | / | 0 | 0.142 |

## B. Band structures of original and Janus monolayers

The band structures of 3 original monolayers and 30 Janus monolayers using PBE method with and without SOC are shown in Fig. S1. When SOC is taken into account, we find that RSS occurs in most Janus monolayers, varying from 0.06 to 0.86 eVÅ. However, for some monolayers, such as Sb$_2$Si$_2$S$_3$Te$_3$, although the RSS also appears at the CBM, it is too weak to be calculated according to the formula, so we do not consider it. In addition, for Sb$_2$SiGeSe$_6$, Bi$_2$SiGeSe$_6$, Sb$_2$Ge$_2$S$_3$Se$_3$, Sb$_2$Si$_2$S$_3$Se$_3$, Bi$_2$Ge$_2$S$_3$Se$_3$, Bi$_2$Si$_2$S$_3$Se$_3$, Sb$_2$SiGeSe$_3$S$_3$, Sb$_2$SiGeS$_3$Se$_3$, Bi$_2$SiGeSe$_3$S$_3$ and Bi$_2$SiGeS$_3$Se$_3$ monolayers, their first valence band (V$_1$) exhibits the Rashba effect at the Γ point, but since it is not at the VBM, the conducting holes or electrons will be disturbed by other states near the Fermi level, so we do not further discuss it.

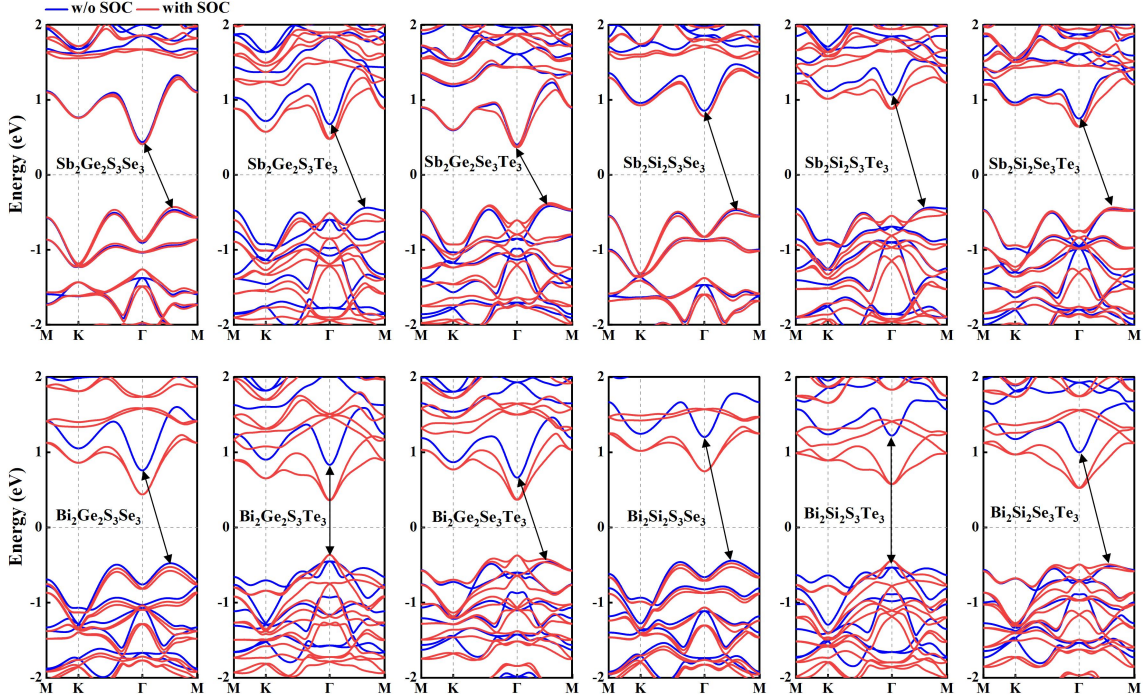

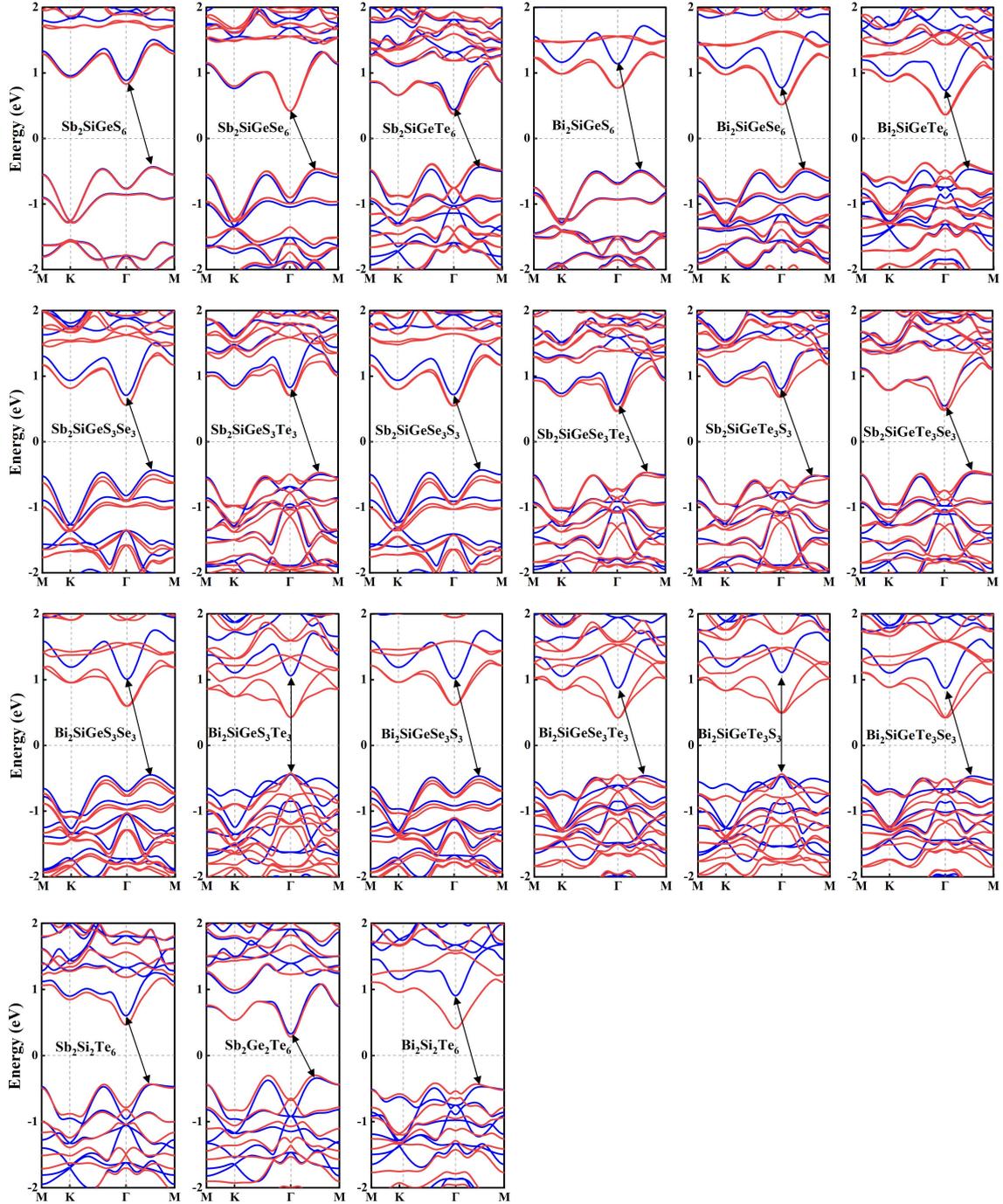

FIG. S1. Band structures of original and Janus monolayers. The bule and red solid line represent the methods of PBE and PBE+SOC, respectively.

## C. Electric field tuning of the Rashba effect

The positive direction of the $E_{ex}$ is the +$z$ direction, and the electric field ranges from -0.3 to 0.3 VÅ$^{-1}$, which can be achieved in the experiment[1]. The calculated variation of the Rashba constant $α_R$ with the $E_{ex}$ is shown in Fig. S2. We find that $α_R$ is linearly dependent on the $E_{ex}$. Furthermore, the electric field response rate（$k\alpha_R = \frac{\Delta\alpha_R}{\Delta E}$）has been calculated, whose $|k\alpha_R|$ values are greater than 0.1 eÅ$^2$.

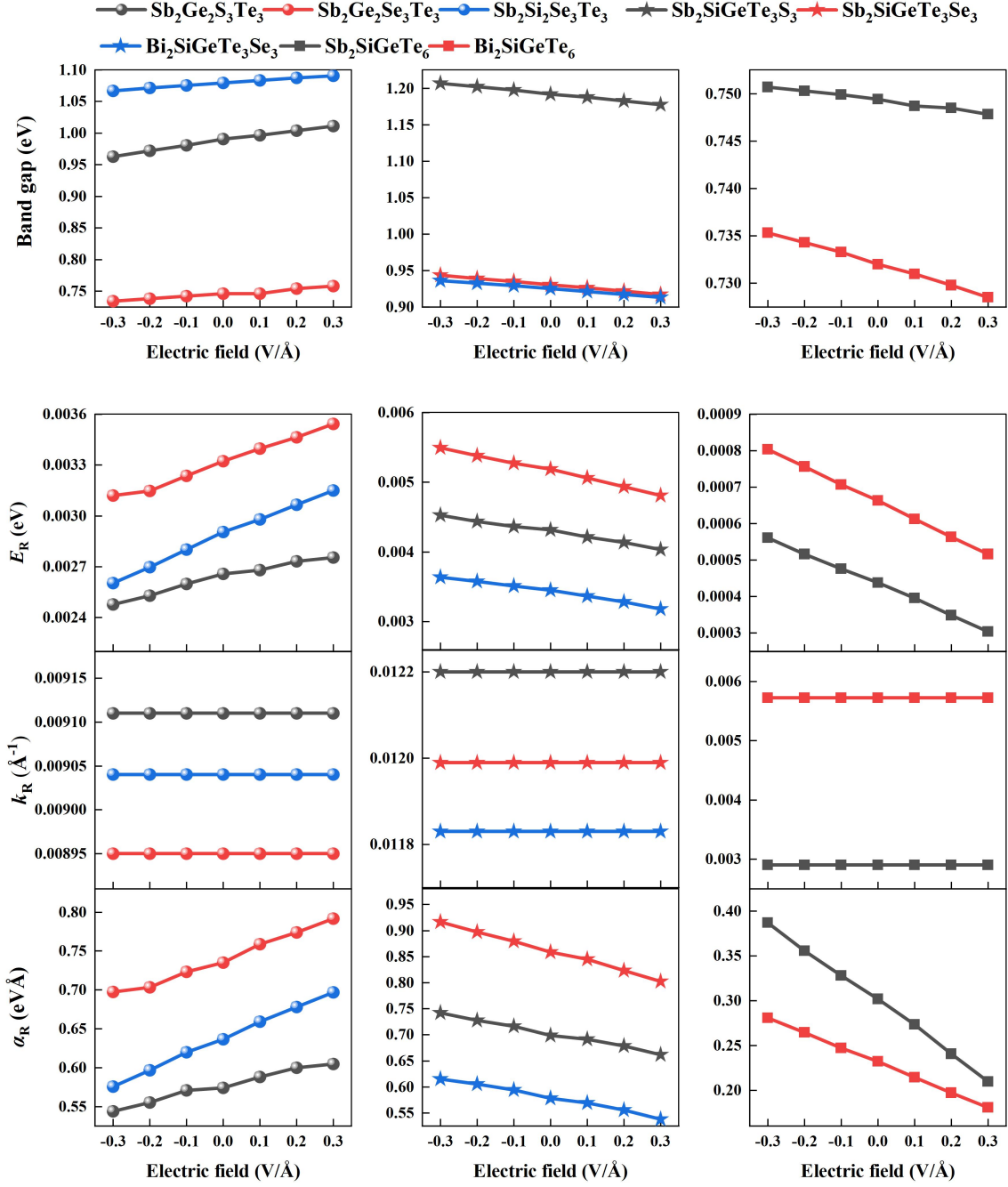

FIG. S2. The band gap, Rashba energy $E_R$, momentum offset $k_R$, Rashba constant $\alpha_R$ of $C_m$-, I-, and $C_p$-Janus monolayers under external electric field.

## D. Polarization properties of Janus monolayers

Previous studies have shown that the breaking of intrinsic mirror symmetry in Janus 2D materials leads to some fascinating polarization-dependent properties such as different surface work functions, dipole moment, and $E_{in}$[2,3]. Specifically speaking, the electrostatic potential energy peaks of the original systems are symmetric, consistent with the symmetric crystal structures, as listed in Fig. S3($a_1$). However, for the Janus systems with broken structural symmetry, the electrostatic potential distributions are asymmetrical with different peaks in the different positions, which leads to different work functions at X and Y atomic surfaces [Fig. S3($a_2$)]. Moreover, we further investigate the mirror symmetry breaking induced dipole moment to confirm the polarization effect (Table SI). The work function difference $\Delta\Phi$ between both surfaces is directly affected by the dipole moment $P$, which is in agreement with the Helmholtz equation. Considering that the vertical intrinsic electric field $E_{in}$ is generally the key factor affecting the $\alpha_R$ of Janus Rashba systems and even its modulation trend under $E_{ex}$, which could be directly reflected in aspects of work function and electrostatic potential, dipole moment, or charge transfer, we have carried out analysis on the above physical quantities (Fig.

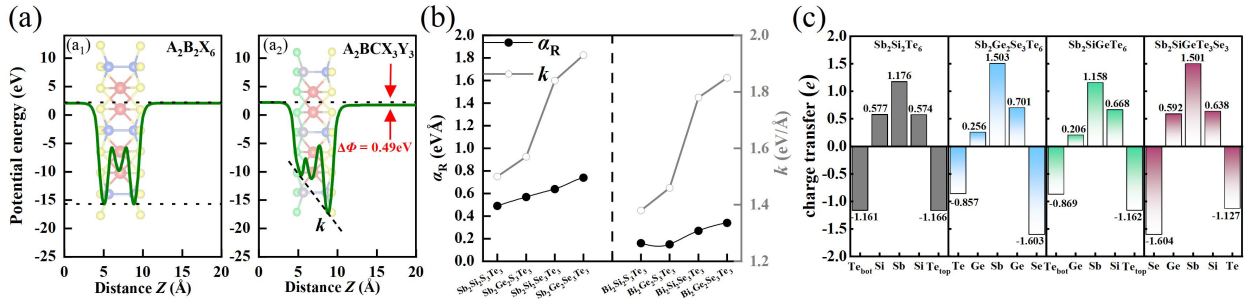

FIG. S3. (a) Planar average of the electrostatic potential of ($a_1$) original and ($a_2$) Janus monolayers along the $z$ direction (taking the original $Sb_2Ge_2Te_6$ and composite Janus $Sb_2SiGeTe_3Se_3$ monolayer as examples). (b) The electrostatic potential slope $k$ on both sides and Rashba constant $\alpha_R$ for the $C_m$-Janus $A_2B_2X_3Y_3$ monolayers. (c) Charge transfer of $Sb_2Si_2Te_6$, $Sb_2Ge_2Se_3Te_3$, $Sb_2SiGeTe_6$, and $Sb_2SiGeTe_3Se_3$ monolayers.

S4).

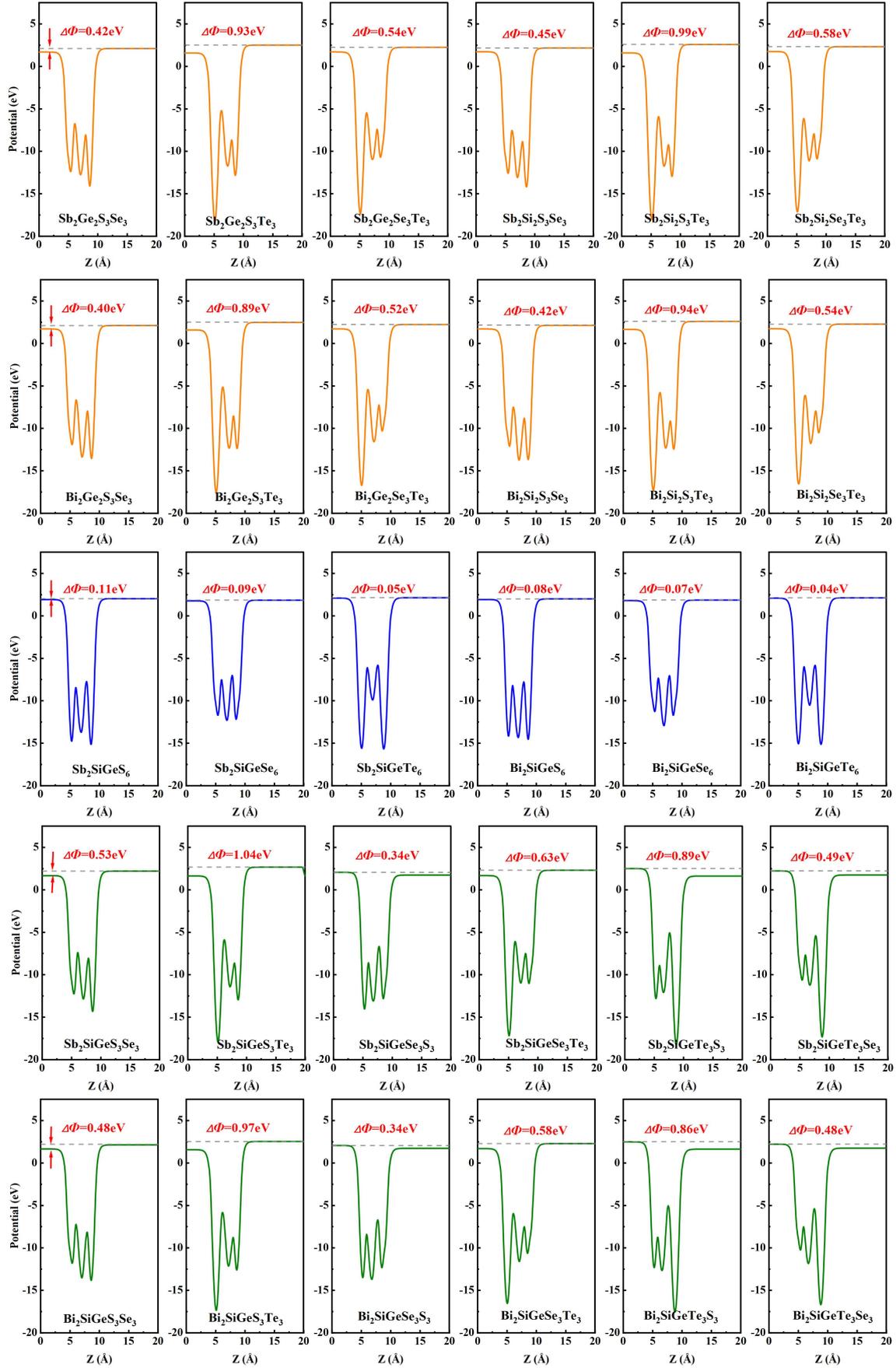

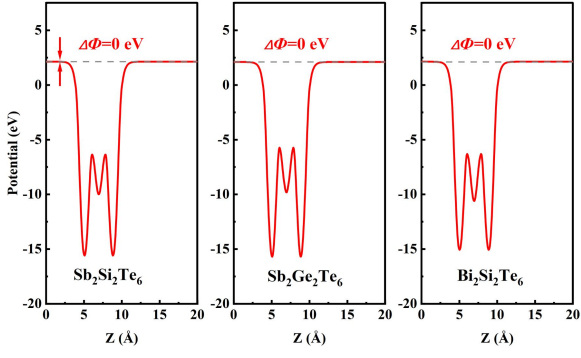

FIG. S4. Planar average of the electrostatic potential of Janus and original monolayers along the $z$ direction. The yellow, blue, green and red solid lines indicate the configurations of $C_m$-, I-, $C_p$-Janus monolayers and original monolayers, respectively.

## E. Atomic adsorption reconfigures charge transfer

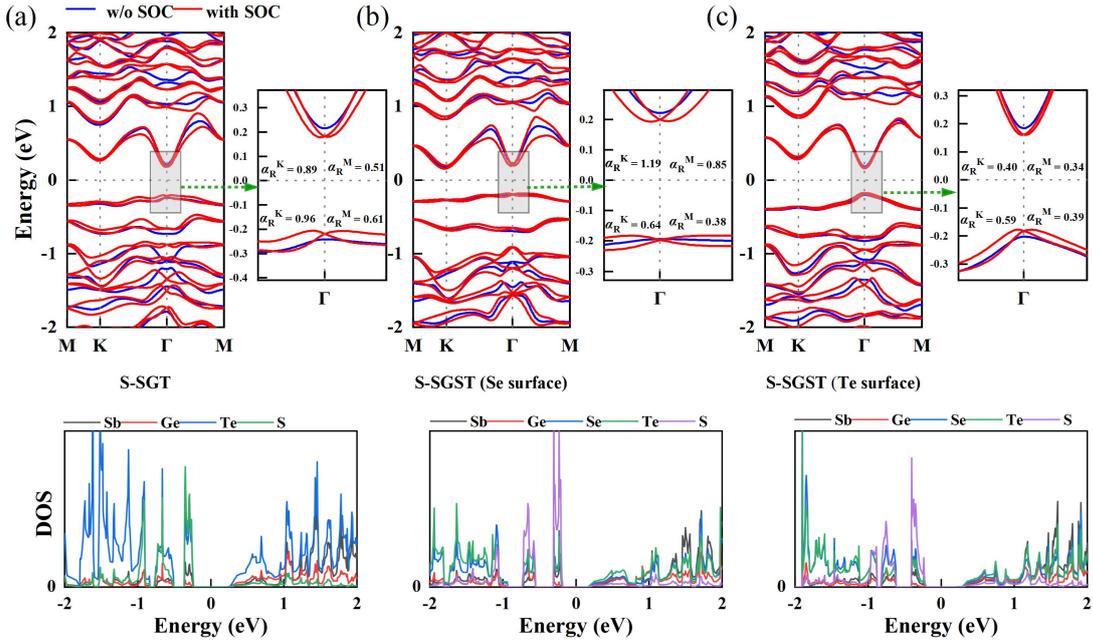

FIG. S5. (a) Band structures and DOS of S adsorbed on SGT. Band structures and DOS of (b) S adsorbed on SGST (Se-surface), and (c) S adsorbed on SGST (Te-surface).

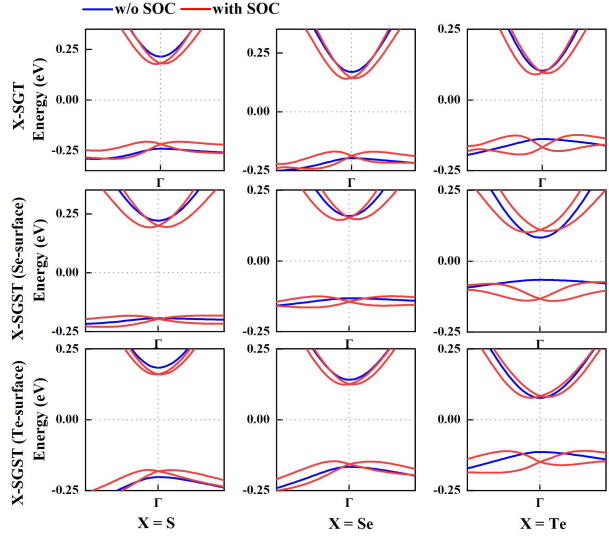

FIG. S6. Band structures of X-SGT and X-SGST systems. Blue and red solid lines indicate methods of PBE and PBE+SOC, respectively.

## F. BN/SGST and BN/S-SGST heterostructures

In experiments, 2D materials are usually grown on a substrate, and the electronic properties are affected by the substrate. Generally, the insulating substrate materials used for 2D devices are BN, AlN, and GaN. The lattice mismatches for SGST/BN (3×3×1), SGST/AlN (2×2×1), and SGST/GaN (2×2×1) are 6%, 12%, and 9%, respectively. Considering practical feasibility and minimizing lattice mismatch between the two stacks, we construct the heterostructure using a 3×3×1 supercell of BN and one SGST layer and investigate their electronic properties. Electronic structure calculations show that heterostructures retain the indirect or direct band gap characteristics of their original systems, but the band gap is significantly reduced compared to the original systems. The band gap of the BN/SGST heterostructures decreases to 0.47 eV (Se-surface) and 0.46 eV (Te-surface), and the band gap of the BN/S-SGST heterostructures decreases to 0.05 eV (Se-surface) and 0.03 eV (Te-surface). When SOC is taken into account, the band gap of the BN/SGST heterostructures decreases to 0.42eV, 0.41 eV, and the band gap of the BN/S-SGST heterostructures decreases to 0.05eV, 0.05 eV. It is worth noting that for the BN/S-SGST system, significant RSS occurs simultaneously at CBM and VBM, which can be attributed to the presence of Te and Se atomic orbitals at CBM and S atomic orbital at VBM, as shown in the projected DOS diagrams of the two heterostructures [Fig. S7(b)].

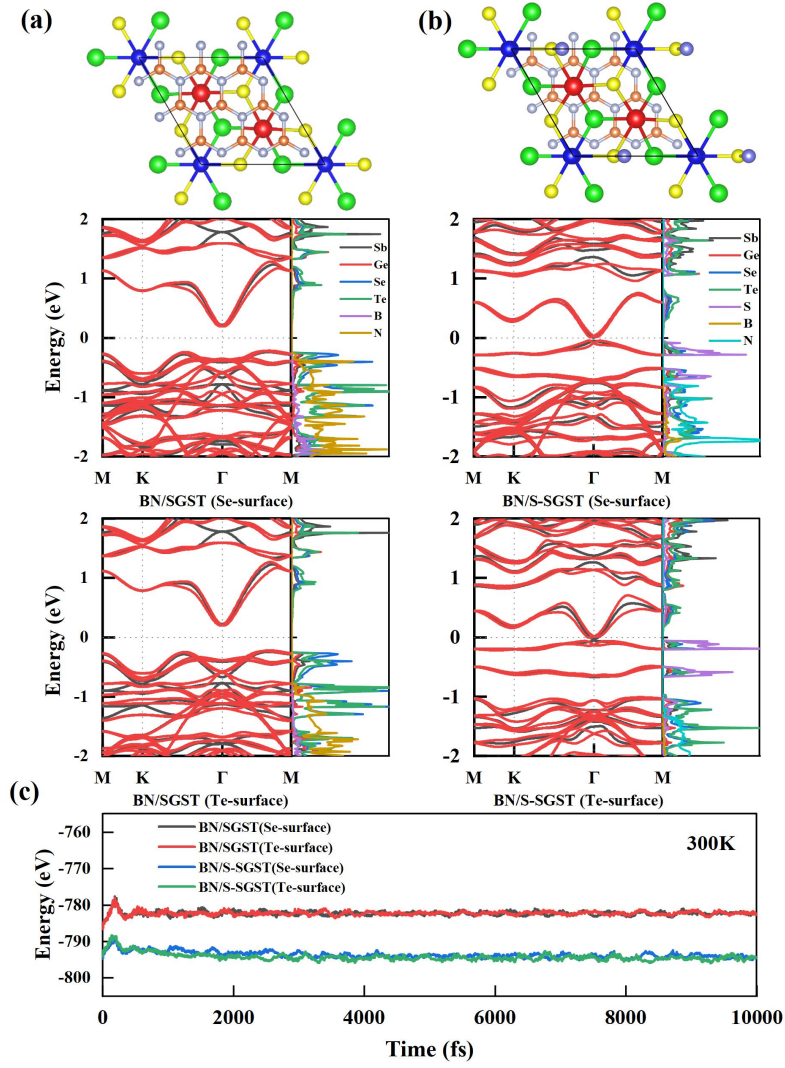

FIG. S7. (a) Band structures and DOS of BN/SGST heterostructures (Se-surface and Te-surface) (b) Band structures and DOS of BN/S-SGST (Se-surface and Te-surface) heterostructures. (c) AIMD simulations of total energy fluctuation under 10 ps at 300K of BN/SGST and BN/S-SGST heterostructures.

# References

1. Domaretskiy, D. *et al.* Quenching the bandgap of two-dimensional semiconductors with a perpendicular electric field. *Nat. Nanotechnol.* **17**, 1078–1083 (2022).
2. Ghobadi, N., Gholami Rudi, S. & Soleimani-Amiri, S. Electronic, spintronic, and piezoelectric properties of new Janus Zn A X Y ( A = Si , Ge , Sn , and X , Y = S , Se , Te ) monolayers. *Phys. Rev. B* **107**, 075443 (2023).
3. Liu, M.-Y., Gong, L., He, Y. & Cao, C. Tuning Rashba effect, band inversion, and spin-charge conversion of Janus X Sn 2 Y monolayers via an external field. *Phys. Rev. B* **103**, 075421 (2021).